# Effects of impurity band on multiphoton photocurrent from InGaN and GaN photodetectors

Chuanliang Wang[1,2], Ahsan Ali[1,2], Jinlei Wu[1,2], Wei Huang[1,2], Hai Lu[3], and Khadga Jung Karki[1,2,4] *

[1]*Department of Physics, Guangdong Technion-Israel Institute of Technology, 241 Daxue Road, Shantou, Guangdong China, 515063;*
[2]*Guangdong Provincial Key Laboratory of Materials and Technologies for Energy Conversion, Guangdong Technion – Israel Institute of Technology, Shantou, Guangdong, China, 515063;*
[3]*School of Electronic Science and Engineering, Collaborative Innovation Center of Advanced Microstructures, Nanjing University, Nanjing 210023, China*
[4]*Technion – Israel Institute of Technology, Haifa, 32000, Israel*

## Abstract

Multiphoton absorption of wide band-gap semiconductors has shown great prospects in many fundamental researches and practical applications. With intensity-modulated femtosecond lasers by acousto-optic frequency shifters, photocurrents and yellow luminescence induced by two-photon absorption of InGaN and GaN photodetectors are investigated experimentally. Photocurrent from InGaN detector shows nearly perfect quadratic dependence on excitation intensity, while that in GaN detector shows cubic and higher order dependence. Yellow luminescence from both detectors show sub-quadratic dependence on excitation intensity. Highly nonlinear photocurrent from GaN is ascribed to absorption of additional photons by long-lived electrons in traps and impurity bands. Our investigation indicates that InGaN can serve as a superior detector for multiphoton absorption, absent of linear and higher order process, while GaN, which suffers from absorption by trapped electrons and impurity bands, must be used with caution.

## Introduction

Gallium Nitride (GaN) is an important wide bandgap material with a variety of applications in LEDs,[1] lasers,[2] power electronics,[3] and UV detectors.[4] It has a bandgap of ~3.4 eV at room temperature. By alloying the material with indium or aluminum, the band gap can be engineered, making it applicable for both visible and UV spectral ranges. Alloys of GaN, mainly InGaN, are also good candidates for photovoltaic devices and visible light communication.[5,6]

Multiphoton absorption of wide bandgap semiconductors like GaN had been reported more than 25 years ago.[7–9] Photocurrents induced by multiphoton absorption have been used to characterize the photodiodes[10,11] and investigate many other ultrafast processes.[12,13] GaN and many other photodiodes have been used as a suitable single-piece replacement for nonlinear optical crystals and photodetector in the characterization of ultrashort pulses.[14,15]

Usually, alloying and doping can improve the optical and electronic properties, thereby pushing the limits of GaN for applications.[16–18] On the other hand, the doping may substantially degrade the performance and reliability of devices substantially.[19–21] Carbon acceptors in GaN are now commonly believed to give rise to the so-called yellow luminescence (YL) near 2.2 eV, which although has been extensively studied still puzzles the scientific community[22–25]. Other impurities, such as Fe and Ca, revealed by spectroscopic measurements also play an important role.[26,27] Recently, YL-related band absorption has been found to be related to photocurrent instability in GaN detectors.[28] The effect of impurities in multiphoton absorption is expected to be more pronounced. GaN detector has been reported to be prone to damage for multiphoton absorption,[29,30] which may be related to the impurities. In this regard, the contributions of alloying/doping and impurity on multiphoton absorption need to be extensively studied to facilitate high-quality nonlinear devices.

In this work, photocurrents and YL induced by multiphoton absorption in GaN and InGaN photodetectors are investigated with phase-modulated femtosecond pulses. Photocurrent in InGaN detector shows nearly perfect quadratic dependence on excitation intensity, while in GaN detector a higher-order intensity dependence is observed. For YL, signals from both detectors show sub-quadratic dependence on excitation intensity. YL-related band absorption is believed to be responsible for the experimental features, which are further verified by lifetime measurement. Our investigation indicates that InGaN is an excellent candidate for multiphoton absorption detector, free of linear and higher-order processes, while GaN is susceptible to higher-order process due to impurity.

## Experimental Setup

The schematic of the experimental setup is shown in Figure 1(A). Mach-Zehnder interferometer is used to generate intensity-modulated femtosecond beams.[31–33] Briefly, a femtosecond laser (Clark MXR Inc., central wavelength at 515 nm, pulse duration of ~200 fs, and repetition rate of 1 MHz) is split into two beams, each of which passes through one of the arms of the interferometer. Acousto-optic frequency shifters (AOFS) in the two arms shift the carrier frequency $v$ to $v_1 = v + p_1$, and $v_2 = v + p_2$, where $p_1$ and $p_2$ are the radio-frequencies (rf) applied to the AOFS. When the two beams are recombined, the output intensity modulates at the difference frequency $\Delta v = p_2 - p_1$, and can be described by $I(t) = \frac{1}{2}I_0(1 + \cos(2\pi \Delta vt))$. $\Delta v$ is set to 1 kHz in our measurements, much less the bandwidth of the devices. According to experimental requirements, either one arm or two arms of Mach-Zehnder interferometer can be adopted, and the single-arm intensity can be modulated by applying a squared wave rf-signal to the AOFS. The output of the interferometer is directed to a Nikon inverted microscope with a 10X objective to excite the sample (GaN or InGaN photodetectors). The back-scattered photoluminescence is separated from the excitation beam by a dichroic mirror, further filtered, and detected by a spectrometer or a photo-multiplier tube (PMT). A current-to-voltage preamplifier is used to convert the photocurrent to voltage, and the resulting electronic signal is digitized by a 24-bit digitizer at the rate of 192k samples per second. The time domain signals are Fourier transformed for further analysis. Photocurrent from a silicon detector is measured to monitor the linearity of electronic components.

InGaN and GaN photodetectors are obtained from *GaNo Optoelectronics Inc*. The detectors are fabricated with *p-i-n* diodes, as shown in Fig. 1(B). For GaN, the diode consists of an Mg-doped *p*-contact layer, an undoped depletion region, a Si-doped *n*-contact layer, and a GaN buffer layer grown on a sapphire substrate by metal–organic chemical vapor deposition (MOCVD). For InGaN detector, the depletion region is made of InGaN/GaN multiple quantum wells instead of undoped GaN. The response curve of two detectors is given in Fig. 1(C). The InGaN detector absorbs below 460 nm and the GaN detector absorbs below 380 nm. The reverse bias is set to 5 V in the measurements.

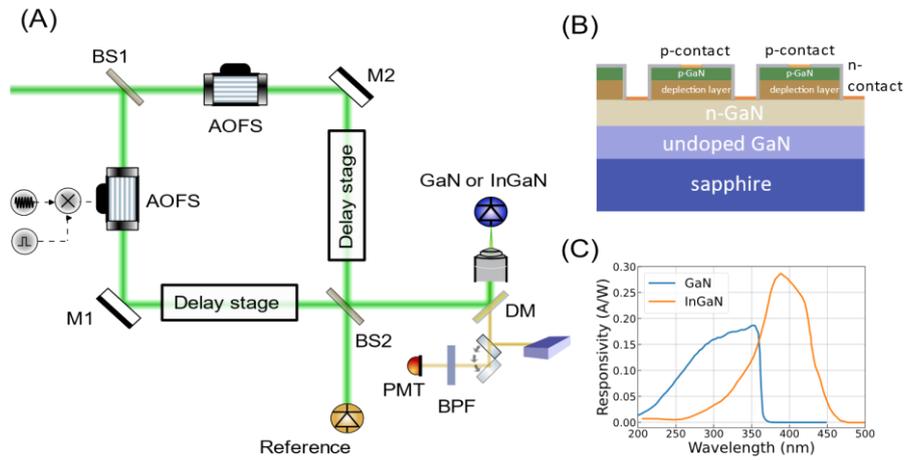

*Figure 1. (A) Schematic of setup. BS: beam splitter, AOFS: acousto-optic frequency shifter, M: mirror, DM: dichroic mirror, BPF: band-pass filter, PMT: photo-multiplier tube. (B) Schematic of InGaN and GaN detector. (C) the response curve of GaN and InGaN detector.*

**Results and Discussion**

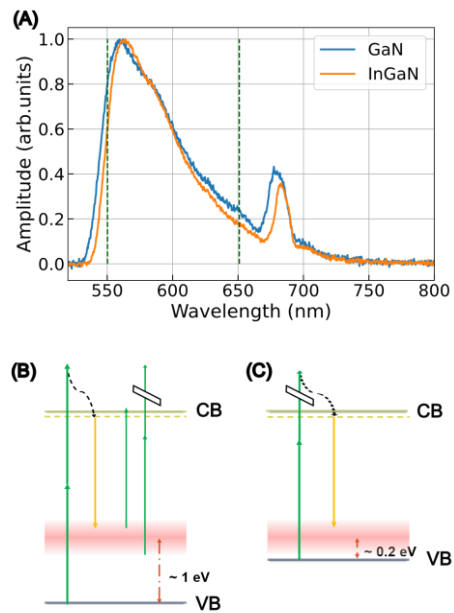

*Figure 2. (A) Normalized emission spectra of GaN and InGaN detector. Emission for wavelength lower than 530 nm was filtered out together with 515 nm excitation light. (B) and (C) shows the band structure of GaN and InGaN, respectively.*

Luminescence is observed both in InGaN and GaN photodetectors when excited with a 515 nm femtosecond pulse. As the energy of a single photon at this wavelength is less than the bandgap of both materials, the excitation is by the absorption of multiple photons. The emission spectra of GaN and InGaN detectors are shown in Fig. 2(A). Both the spectra are similar. They span from ~530 nm to ~700 nm, and a sharp peak is located at ~680 nm. The sharp peak is usually attributed to $Cr^{3+}$ excitation in the sapphire substrate.[24,34] The broad emission spectrum is YL, which is due to the impurity band of $C_N$ acceptors.[22,23] Overall spectra of YL from multiphoton absorption are similar to the ones from single-photon absorption.[35,36] In functional devices like UV detectors and LEDs, near-band-edge emission is concomitant and usually used to characterize the diode,[37,38] while here the presence of YL provides a signal that can be used to check how the impurity band affects the functionality of photodiodes.

Both the materials, InGaN and GaN, have a direct bandgap that can be tuned by the composition of In and Ga within the alloy. In our case, the bandgaps are ~2.6 eV and ~3.4 eV for InGaN and GaN detectors, respectively.

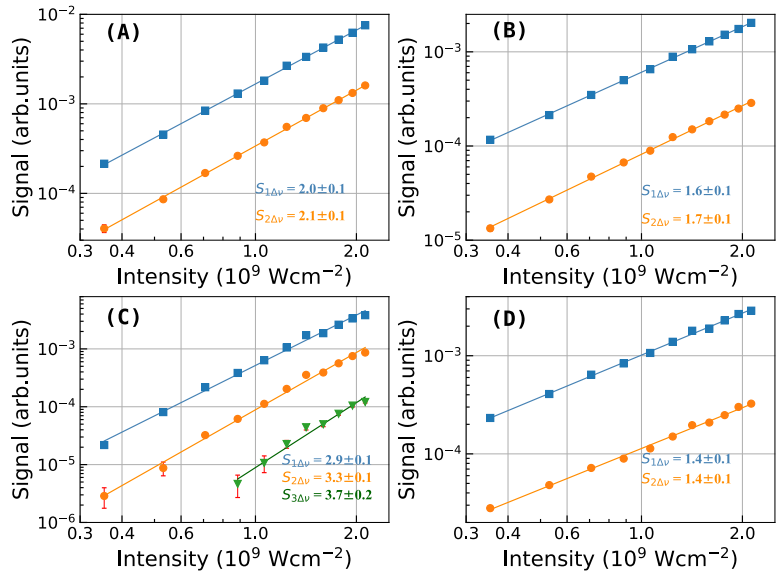

*Figure 3. Intensity dependence of signals for (A) Photocurrent of InGaN, (B) YL of InGaN, (C) Photocurrent of GaN, and (D) YL of GaN subjected to intensity-modulated femtosecond pulses at 515 nm.*

In order to investigate the role of the impurity band on the nonlinear absorption in InGaN and GaN, we measure the intensity dependence of photocurrents and YL excited by an intensity-modulated beam of femtosecond laser pulses. Here, we mainly focus on the impurity band of $C_N$ acceptors. The emission at 680 nm is filtered out from the YL by suitable filters. Only the photons in the spectra range of 550 -- 650 nm are detected by PMT. Intensity modulation by Mach-Zehnder interferometer with AOFS provides a simple method to distinguish signals from different orders of nonlinearity.[33] If the intensity of the excitation beam is modulated at a single frequency $\Delta v$, the signals, such as photocurrent and luminescence, due to two-photon absorption (TPA) modulates at $2\Delta v$ as well as $1\Delta v$. The strength of both signals show quadratic dependence on the intensity of modulation.[32,39] If the signals arise from different absorption pathways, for example, TPA as well as single photon absorption, the signal at $\Delta v$ shows sub-quadratic intensity dependence, while the signal at $2\Delta v$ still shows a quadratic dependence. Similarly, if three-photon absorption contributes to the signal then an additional peak at $3\Delta v$ emerges with a cubic intensity dependence.

The strength of photocurrent from InGaN at the modulation frequencies of $\Delta v$ and $2\Delta v$ are shown in Fig. 3(A). Both the signals show quadratic dependence on the intensity of the excitation beam indicating the prevalence of TPA in the photocurrent, which is expected in a pure semiconductor. This is surprising because YL from the system clearly shows the presence of impurities, and it raises the question of the role of impurities on the photocurrent in the device. Can the impurity band be populated by the absorption of a single sub-bandgap photon? If yes, what is the effect of such process on the photocurrent? These are important questions because in some wide-bandgap semiconductors, the photocurrent signal from the sub-bandgap absorption is comparable to that from TPA.[33]

Next, we analyze the YL from the InGaN photodetector to investigate how the impurity band affects functionality. The dependence of YL on the excitation intensity is shown in Fig. 3(B). Unlike photocurrent, YL does not increase quadratically with the excitation as inferred from the fits of the log-log plots. The slopes of the signals at $1\Delta v$ and $2\Delta v$ are 1.6 ± 0.1 and 1.7 ± 0.2, respectively. The fact that the slopes in YL and photocurrent are different indicates that the two signals arise from different dynamical evolution of the excitation. The sub-quadratic slopes in YL have two possible causes: i) the impurities can be populated by single as well as TPA or ii) these bands are populated by the relaxation of electrons from the conduction band wherein the population dynamics also affects the dependence on the excitation. The former case can be ruled out as the emission has been shown to arise due to the relaxation of electrons from the conduction band to the impurity band.[22] In the latter case, saturation of the impurity band can lead to the power law with the exponent less than 2 for TPA. These effects have been observed previously under single and multiple photon absorption in GaN.[25,36,40]

Now we analyze photocurrent and YL from GaN from comparison. The dependence of the modulated photocurrent signals on excitation intensity is shown in Fig. 3(C). In addition to $\Delta v$ and $2\Delta v$, the modulation is present at $3\Delta v$. The slopes of the signals in the log−log plots are 2.9±0.1, 3.3±0.1 and 3.7±0.2, respectively. As the exponent of the power law is more than 2, photocurrent in GaN cannot be explained by a simple TPA. YL, on the other hand, is modulated only at $\Delta v$ and $2\Delta v$ (see Fig. 3(D)). Although the exponent of the power law for the two signals is 1.4±0.1, which is less than that in InGaN, it can still be attributed to TPA and saturation of the impurity bands. While the modulation at $\Delta v$ and $2\Delta v$ and the sub-quadratic increase with excitation intensity in YL from GaN is of similar origin as from InGaN, the modulation of photocurrent from GaN at $3\Delta v$ indicates that the physical processes that determine the functionality of the two systems are not identical.

There are two possible origins of the photocurrent signal at $3\Delta v$: i) Excitation of electrons to high-lying bands by three or four-photon absorption, or ii) relaxation of electrons from the conduction band to a long-lived trap state from which they can be re-excited by absorption of additional photons from subsequent laser pulses. The first case is unlikely as the similarity in the band structure of InGaN and GaN implies that higher-order multiphoton absorption should also be evident in InGaN, which is not observed in our measurements. The second case posits long-lived trap states that can be confirmed by measurements of emission kinetics.

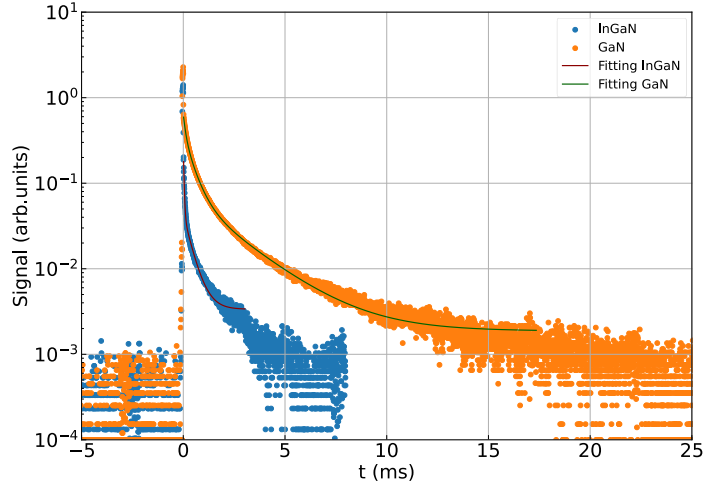

*Figure 4. (A) Lifetime of YL from InGaN and GaN photodetectors. The curves are averaged over $10^5$ gated squares. The experimental data are fitted with multiple exponential decays $S(t) = \sum A_i e^{-t/\tau_i} + c$.*

Fig. 4 shows the time-resolved YL from both samples. In order to measure the kinetics, the excitation beam was switched on and off by applying a 20 Hz square wave rf signal with a duty cycle of 0.1% (*i. e.* 50 pulses) to one of the AOFS. The emissions show multiple exponential decay. The strength and lifetimes of the decay processes are given in Table 1.

|       | $A_1$ | $\tau_1$ (µs) | $A_2$ | $\tau_2$ (µs) | $A_3$ | $\tau_3$ (µs) |
|-------|-------|---------------|-------|---------------|-------|---------------|
| InGaN | 0.21  | 35            | 0.04  | 420           | --    | --            |
| GaN   | 0.26  | 110           | 0.3   | 460           | $7.2 \times 10^{-2}$ | 2300 |

*Table 1 Fitting parameters for decay curve of YL from InGaN and GaN. YL from InGaN shows biexponential decay, while YL from GaN shows a third exponential decay extending to milliseconds.*

The fast component of the decay is a convolution of the response function of the measurement system and radiative relaxation from the conduction band to the $C_N$ impurity band. The slower decays are from the radiative relaxation of trapped electrons to the impurity band. Note that our measurements agree with previous observations of YL with a long lifetime.[24,41] Such decays typically have a distribution of lifetimes. Here, we have used the lowest number of decay components to fit the kinetics. In InGaN, we only need an additional slow decay with a time constant of 420 µs, while in GaN we need two additional components with time constants of 460 μs and 2.3 ms. The faster component is similar to that in InGaN. The slower component indicates the presence of very long-lived traps that are more significant in the YL from GaN. Now, in the measurements with modulated femtosecond pulses, subsequent pulses arrive before the electrons relax to the valence band leading to the re-excitation of electrons from the impurity band or the traps. This is a 2 + 1 (or even 2) resonant-like process, for which the intensity dependence is higher than quadratic, and which also modulates the resulting photocurrent at 3$\Delta v$. Clearly, the effects of long-lived traps and impurity bands are more pronounced in GaN than in InGaN. To understand why this is the case, we note that dopants like Si and Ge are known to significantly shorten the lifetime of YL in GaN.[24] The same could be true with indium alloying.

## CONCLUSION

In this work, photocurrents and yellow luminescence induced by two-photon absorption for InGaN and GaN photodetectors are investigated with phase-modulated femtosecond lasers. Photocurrents from InGaN detector show nearly perfect quadratic dependence on laser intensity, while those for GaN detector exhibit cubic and higher intensity dependence. YL signals from both detectors show sub-quadratic dependence on excitation intensity, but the deviation for GaN is larger. Measurement of lifetime of YL shows that a significant fraction of emission occurs at a timescale longer than a microsecond in GaN, which points to the accumulation of electrons in the long-lived traps. Reabsorption from traps and impurity bands contributes to photocurrent resulting in a highly nonlinear response from the GaN photodetectors. Alloying with In significantly reduces the lifetime of the traps and thereby accumulation of electrons in them. Thus, InGaN behaves close to an ideal two-photon detector. These findings are important in designing nonlinear optoelectronic components from wide bandgap semiconductors for high-frequency applications.

## ACKNOWLEDGMENTS

This work was funded by the Guangdong Province Science and Technology Major Project: Future functional materials under extreme conditions (FMUXC)----

**REFERENCE**

[1] S. Nakamura, and M.R. Krames, "History of gallium-nitride-based light-emitting diodes for illumination," Proc. IEEE **101**(10), 2211–2220 (2013).

[2] J.C. Johnson, H.J. Choi, K.P. Knutsen, R.D. Schaller, P. Yang, and R.J. Saykally, "Single gallium nitride nanowire lasers," Nat. Mater. **1**(2), 106–110 (2002).

[3] M. Meneghini, C. De Santi, I. Abid, M. Buffolo, M. Cioni, R.A. Khadar, L. Nela, N. Zagni, A. Chini, F. Medjdoub, G. Meneghesso, G. Verzellesi, E. Zanoni, and E. Matioli, "GaN-based power devices: Physics, reliability, and perspectives," J. Appl. Phys. **130**(18), 1–83 (2021).

[4] H.D. Jabbar, M.A. Fakhri, and M. Jalal Abdulrazzaq, "Gallium Nitride -Based Photodiode: A review," Mater. Today Proc. **42**, 2829–2834 (2021).

[5] D.V.P. McLaughlin, and J.M. Pearce, "Progress in indium gallium nitride materials for solar photovoltaic energy conversion," Metall. Mater. Trans. A **44**(4), 1947–1954 (2013).

[6] D. Kong, Y. Zhou, J. Chai, S. Chen, L. Chen, L. Li, T. Lin, W. Wang, and G. Li, "Recent progress in InGaN-based photodetectors for visible light communication," J. Mater. Chem. C, 14080–14090 (2022).

[7] J. Miragliotta, and D.K. Wickenden, "Transient photocurrent induced in gallium nitride by two-photon absorption," Appl. Phys. Lett. **69**(14), 2095–2097 (1996).

[8] S. Petit, D. Guennani, P. Gilliot, C. Hirlimann, B. Hönerlage, O. Briot, and R.L. Aulombard, "Luminescence and absorption of GaN films under high excitation," Mater. Sci. Eng. B **43**, 196–200 (1997).

[9] D. Kim, I. Libon, C. Voelkmann, Y. Shen, and V. Petrova-Koch, "Multiphoton photoluminescence from GaN with tunable picosecond pulses," Phys. Rev. B **55**(8), R4907 (1997).

[10] C. Xu, and W. Denk, "Two-photon optical beam induced current (OBIC) imaging through the backside of integrated circuits," Appl. Phys. Lett. **71**, 2578–2580 (1997).

[11] F. Kao, M. Huang, Y. Wang, M. Lee, and C.-K. Sun, "Two-photon optical-beam-induced current imaging of indium gallium nitride blue light-emitting diodes," Opt. Lett. **24**(20), 1407–1409 (1999).

[12] Q. Bian, F. Ma, S. Chen, Q. Wei, X. Su, I.A. Buyanova, W.M. Chen, C.S. Ponseca, M. Linares, K.J. Karki, A. Yartsev, and O. Inganäs, "Vibronic coherence contributes to photocurrent generation in organic semiconductor heterojunction diodes," Nat. Commun. **11**, 617 (2020).

[13] D.A. Fishman, C.M. Cirloganu, S. Webster, L.A. Padilha, M. Monroe, D.J. Hagan, and E.W. Van Stryland, "Sensitive mid-infrared detection in wide-bandgap semiconductors using extreme non-degenerate two-photon absorption," Nat. Photonics **5**(9), 561–565 (2011).

[14] A.M. Streltsov, K.D. Moll, A.L. Gaeta, P. Kung, D. Walker, and M. Razeghi, "Pulse autocorrelation measurements based on two- and three-photon conductivity in a GaN photodiode," Appl. Phys. Lett. **75**(24), 3778–3780 (1999).

[15] M. Zürch, A. Hoffmann, M. Gräfe, B. Landgraf, M. Riediger, and C. Spielmann, "Characterization of a broadband interferometric autocorrelator for visible light with ultrashort blue laser pulses," Opt. Commun. **321**, 28–31 (2014).

[16] M.G. Vivas, D.S. Manoel, J. Dipold, R.J. Martins, R.D. Fonseca, I. Manglano-Clavero, C. Margenfeld, A. Waag, T. Voss, and C.R. Mendonca, "Femtosecond-laser induced two-photon absorption of GaN and $Al_xGa_{1-x}N$ thin films: Tuning the nonlinear optical response by alloying and doping," J. Alloys Compd. **825**, 153828 (2020).

[17] M. Hetzl, M. Kraut, J. Winnerl, L. Francaviglia, M. Döblinger, S. Matich, A. FontcubertaMorral, and M. Stutzmann, "Strain-Induced Band Gap Engineering in Selectively Grown GaN-(Al,Ga)N Core-Shell Nanowire Heterostructures," Nano Lett. **16**(11), 7098–7106 (2016).

[18] S. Zhou, H. Xu, H. Hu, C. Gui, and S. Liu, "High quality GaN buffer layer by isoelectronic doping and its application to 365 nm InGaN/AlGaN ultraviolet light-emitting diodes," Appl. Surf. Sci. **471**, 231–238 (2019).

[19] M. Kato, T. Asada, T. Maeda, K. Ito, K. Tomita, T. Narita, and T. Kachi, "Contribution of the carbon-originated hole trap to slow decays of photoluminescence and photoconductivity in homoepitaxial n-type GaN layers," J. Appl. Phys. **129**(11), 115701 (2021).

[20] C. Haller, J.F. Carlin, G. Jacopin, W. Liu, D. Martin, R. Butté, and N. Grandjean, "GaN surface as the source of non-radiative defects in InGaN/GaN quantum wells," Appl. Phys. Lett. **113**(11), 111106 (2018).

[21] R.M. Barrett, J.M. McMahon, R. Ahumada-Lazo, J.A. Alanis, P. Parkinson, S. Schulz, M.J. Kappers, R.A. Oliver, and D. Binks, "Disentangling the Impact of Point Defect Density and Carrier Localization-Enhanced Auger Recombination on Efficiency Droop in (In,Ga)N/GaN Quantum Wells," ACS Photonics **10**, 2632–2640 (2023).

[22] F. Zimmermann, J. Beyer, C. Röder, F.C. Beyer, E. Richter, K. Irmscher, and J. Heitmann, "Current Status of Carbon-Related Defect Luminescence in GaN," Phys. Status Solidi **218**(20), 2100235 (2021).

[23] M.A. Reshchikov, "On the Origin of the Yellow Luminescence Band in GaN," Phys. Status Solidi **2200488**, (2022).

[24] T. Vanek, V. Jary, T. Hubacek, F. Hajek, K. Kuldova, Z. Gedeonova, V. Babin, Z. Remes, and M. Buryi, "Acceleration of the yellow band luminescence in GaN layers via Si and Ge doping," J. Alloys Compd. **914**, 165255 (2022).

[25] Y. Toda, T. Matsubara, R. Morita, M. Yamashita, K. Hoshino, T. Someya, and Y. Arakawa, "Two-photon absorption and multiphoton-induced photoluminescence of bulk GaN excited below the middle of the band gap," Appl. Phys. Lett. **82**(26), 4714–4716 (2003).

[26] M.A. Reshchikov, D.O. Demchenko, M. Vorobiov, O. Andrieiev, B. McEwen, F. Shahedipour-Sandvik, K. Sierakowski, P. Jaroszynski, and M. Bockowski, "Photoluminescence related to Ca in GaN," Phys. Rev. B **106**, 035206 (2022).

[27] J. Wang, F. Shi, X. Wu, J. Yang, Y. Chen, Q. Wu, Y. Song, and Y. Fang, "Ultrafast broadband carrier and exciton dynamics of Fe-related centers in GaN," Appl. Phys. Lett. **123**, 042107 (2023).

[28] L. Pu, Z. Wang, D. Zhou, W. Xu, F. Ren, D. Chen, R. Zhang, Y. Zheng, and H. Lu, "Yellow luminescence band defect related photocurrent instability of GaN p-i-n ultraviolet photodetectors," J. Vac. Sci. Technol. B **40**(5), 052201 (2022).

[29] M. Tripepi, S. Zhang, B. Harris, N. Talisa, J.H. Yoo, H. Peelaers, S. Elhadj, and E. Chowdhury, "Few-cycle optical field breakdown and damage of gallium oxide and gallium nitride," APL Mater. **10**(7), 071107 (2022).

[30] A. Bourgine, D. Lagarde, S. Dubos, J. Guillermin, N. Chatry, C. Chatry, M. Mauguet, and X. Marie, "Multiphoton Absorption in Gallium Nitride and Silicon Carbide Photodiodes: Applications for Single Event Effects Tests," IEEE Trans. Nucl. Sci. **70**(7), 1451–1458 (2023).

[31] P. Kumar, and K.J. Karki, "Two-photon excitation spectroscopy of 1,5–diphenyl-1,3,5-hexatriene using phase modulation," J. Phys. Commun. **3**(3), 35008 (2019).

[32] K.J. Karki, L. Kringle, A.H. Marcus, and T. Pullerits, "Phase-synchronous detection of coherent and incoherent nonlinear signals," J. Opt. **18**, 015504 (2015).

[33] C. Wang, J. Cai, X. Liu, C. Chen, X. Chen, and K.J. Karki, "In Operando Quantification of Single and Multiphoton Photocurrents in GaP and InGaN Photodetectors with Phase-Modulated Femtosecond Light Pulses," ACS Photonics **10**, 1119–1125 (2023).

[34] H.H. Kusuma, B. Astuti, and Z. Ibrahim, "Absorption and emission properties of ruby (Cr:Al2O3) single crystal," J. Phys. Conf. Ser. **1170**(1), 012054 (2019).

[35] R.H. Godiksen, T.S. Aunsborg, P.K. Kristensen, and K. Pedersen, "Two-photon photoluminescence and second-harmonic generation from unintentionally doped and semi-insulating GaN crystals," Appl. Phys. B **123**, 270 (2017).

[36] R.J. Martins, J.P. Siqueira, I. Manglano Clavero, C. Margenfeld, S. Fündling, A. Vogt, A. Waag, T. Voss, and C.R. Mendonca, "Carrier dynamics and optical nonlinearities in a GaN epitaxial thin film under three-photon absorption," J. Appl. Phys. **123**, 243101 (2018).

[37] H. Yang, Z. Ma, Y. Jiang, H. Wu, P. Zuo, B. Zhao, H. Jia, and H. Chen, "The enhanced photo absorption and carrier transportation of InGaN/GaN Quantum Wells for photodiode detector applications," Sci. Rep. **7**, 43357 (2017).

[38] M. Meneghini, S. Vaccari, N. Trivellin, D. Zhu, C. Humphreys, R. Butendheich, C. Leirer, B. Hahn, G. Meneghesso, and E. Zanoni, "Analysis of defect-related localized emission processes in InGaN/GaN-based LEDs," IEEE Trans. Electron Devices **59**(5), 1416–1422 (2012).

[39] P. Tian, and W.S. Warren, "Ultrafast measurement of two-photon absorption by loss modulation," Opt. Lett. **27**(18), 1634–1636 (2002).

[40] W. Grieshaber, E.F. Schubert, I.D. Goepfert, R.F. Karlicek, M.J. Schuman, and C. Tran, "Competition between band gap and yellow luminescence in GaN and its relevance for optoelectronic devices," J. Appl. Phys. **80**(8), 4615–4620 (1996).

[41] M.A. Reshchikov, J.D. McNamara, H. Helava, A. Usikov, and Y. Makarov, "Two yellow luminescence bands in undoped GaN," Sci. Rep. **8**, 8091 (2018).